\documentclass[aps,twocolumn,showpacs,pre,superscriptaddress]{revtex4-1}

\usepackage{graphicx}
\usepackage{amsmath}
\usepackage{amssymb}
\usepackage{bm}
\usepackage{color}
\usepackage{ulem}

\begin{document}

\title{Housekeeping Entropy in Continuous Stochastic Dynamics with Odd-Parity Variables}

\author{Joonhyun~Yeo}
\affiliation{
School of Physics, Konkuk University, Seoul 143-701, Korea}
\author{Chulan~Kwon}
\affiliation{Department of Physics, Myongji University, Yongin,
Gyeonggi-Do 449-728, Korea}
\author{Hyun Keun Lee}
\affiliation{Department of Physics and Astronomy, Seoul National University,
Seoul 151-747, Korea}
\author{Hyunggyu~Park}
\affiliation{School of Physics, Korea Institute for
Advanced Study, Seoul 130-722, Korea}

\date{\today}

\begin{abstract}
We investigate the decomposition of the total entropy production
in continuous stochastic dynamics when there are
odd-parity variables that change their signs under time reversal.
The first component of the entropy production, which
satisfies the fluctuation theorem, is associated with the usual excess heat
that appears during transitions between stationary states.
The remaining housekeeping part of the entropy production can be further split into two parts.
We show that this decomposition can be achieved in infinitely many ways characterized by a single parameter $\sigma$.
For an arbitrary value of $\sigma$, one of the two parts contributing
to the housekeeping entropy production satisfies the fluctuation theorem.
We show that for a range of $\sigma$ values this part can be associated with
the breakage of the detailed balance in the steady state, and can be regarded as
a continuous version of the corresponding entropy production that has been obtained
previously for discrete state variables.
The other part of the housekeeping entropy does not satisfy the fluctuation theorem
and is related to the parity asymmetry of the stationary state
distribution. We discuss our results in connection with the difference between continuous and discrete variable cases
especially in the conditions for the detailed balance and the parity symmetry of the stationary state distribution.
\end{abstract}



\maketitle

\section{Introduction}

Recent advances in understanding nonequilibrium systems have been stimulated by the discovery and applications of the
so-called fluctuation theorems (FTs)~\cite{evans,gallavotti,jarzynski,crooks,kurchan,lebowitz}.
 The total entropy production (EP) in the system and the heat reservoir plays a central role as a measure of the irreversibility in the nonequilibrium dynamics.
The inequality for the total EP in the thermodynamic second law now becomes a corollary of
the more general equality in the FTs \cite{crooks,jarzynski,kurchan}.
In a further development, a part of the total EP associated with the excess heat is shown to satisfy a
FT on its own \cite{hatano}. The remaining part is the housekeeping EP that is necessary
to maintain the nonequilibrium steady state. It was also shown to satisfy the separate FT \cite{speck,seifert}.
The excess entropy is a transient component of the total EP as it is
produced during transitions between stationary states. It can therefore be interpreted as
a nonadiabatic component of the total EP, while the adiabatic part corresponds
to the housekeeping EP \cite{esposito}.

The situations become more complicated, however, when there are dynamical variables
which have an odd-parity under time reversal.
An example of odd-parity variables is the momentum in the underdamped description of the Brownian motion.
In a series of papers \cite{spinney,spinney1}, Spinney and Ford considered the EP
in both continuous and discrete stochastic systems in the presence of
odd-parity variables. They found that the excess entropy part can still be singled out, which satisfies the FT.
The housekeeping part is further divided into two terms $\Delta S_2$ and $\Delta S_3$.
Only $\Delta S_2$ is shown to satisfy the FT.  $\Delta S_3$ is associated with the asymmetry of the
steady-state distribution under time reversal. It does not satisfy the FT and
turns out to be transient as it vanishes when the system stays in a steady state.
In a more recent study \cite{lkp} on a stochastic system described by discrete variables including odd-parity ones,
the housekeeping EP was found to be decomposed into $\Delta S_{\rm bDB}$ and $\Delta S_{\rm as}$ in a manner different
from that used by  Spinney and Ford. $\Delta S_{\rm bDB}$ is a direct consequence
of the breakage of the detailed balance (DB) and satisfies the FT.
$\Delta S_{\rm as}$ is attributed to the asymmetry of the steady-state distribution as in the Spinney and Ford's scheme,
but turns out to be not transient in this case. It can therefore be regarded as
a component of the adiabatic EP. For continuous variables, we note that
it is not obvious how $\Delta S_2$ found by Spinney and Ford \cite{spinney}
is related to the breakage of DB.

It is the purpose of this paper to identify a part of the housekeeping EP as a proper measure of
the DB breakage as done in the case of discrete state variables.
We consider a general continuous stochastic system with odd-parity variables
described by the Fokker-Planck equation including the driven Brownian motion
in the underdamped limit, where the momentum serves as the odd-parity variable.
As we shall see below, there is a  subtle difference between
the discrete-jumping process described by the master equation and
the continuous-evolving process described by the Fokker-Planck equation.
As a result, the direct application of the procedures used in Ref.~\cite{lkp}
for obtaining $\Delta S_{\rm bDB}$ to the continuous system makes the corresponding quantity ill-defined.
One has to resort to a different approach to obtain a sensible result.
We note that a convenient way to obtain various EPs is to use a dual or adjoint dynamics to the
original dynamics \cite{chernyak,seifert_review}. In this paper, we find that many versions of
$\Delta S_{\rm bDB}$ for continuous variables can be obtained via a generalized adjoint
dynamics like the one used in Ref.~\cite{gen_adj} where an infinite family of
the excess EPs are constructed.
We will show below that, in the housekeeping EP, we can identify infinitely many expressions for the EP related to the DB breakage, $\Delta S_{\rm bDB}^\sigma$, parametrized by an arbitrary number $\sigma$,
all of which satisfy the FT. Spinney and Ford's $\Delta S_2$
corresponds to a special case of $\sigma=0$.

In the next section, we present how various components of the EP are defined in the general continuous stochastic dynamics
including the driven Brownian dynamics. The explicit expressions for the total and the excess
EPs are given along with their average rates in time.
Although much of the results in the section are previously known, we present them to
clear up certain technical points and to set up our notations.
In Sec.~\ref{sec:hk}, we present our main results.
We show in detail how the housekeeping EP can be divided
into two parts by applying the generalized
adjoint dynamics. We study in detail how the DB and its breakage can be represented in
the expressions of the EPs.
In the final section, we summarize our results with discussion.

\section{Entropy Productions in Continuous Stochastic Dynamics}

A simple model for the continuous stochastic dynamics involving odd-parity variables is
the driven Brownian dynamics for a particle of mass $m$ in $d$ dimensions, for which
the position and momentum variables, $\bm{q}=(\bm{x},\bm{p})$,
constitute the even and odd parity variables, respectively.
The equations of motion are given by
\begin{eqnarray}
&&\dot{\bm{x}}=\frac{\bm{p}}{m}, \label{x_dot}\\
&&\dot{\bm{p}}=-\mathsf{G}\cdot\frac{\bm{p}}{m}+\bm{f}(\bm{q};\lambda)
+\boldsymbol{\xi}(t),
\end{eqnarray}
where $\mathsf{G}=\{G_{ij}\}$ is a dissipation matrix with the 
standard notation $(\mathsf{G}\cdot\bm{p})_i=\sum_j G_{ij}p_j$. We consider a most general form
for the force which may depend on
both position and momentum as well as on some time-dependent protocol $\lambda(t)$.
In the following, we will drop the expression for $\lambda$-dependence in the force for simplicity of notation.
The Gaussian white noise satisfies
\begin{eqnarray}
\langle\boldsymbol{\xi}(t)\rangle=0,~~~~~
\langle \xi_i(t)\xi_j(t^\prime)\rangle=2D_{ij}\delta(t-t^\prime)
\label{noise}
\end{eqnarray}
with the symmetric positive-definite diffusion matrix $\mathsf{D}=\{D_{ij}\}$. If the Einstein relation holds, $\mathsf{D}=\mathsf{G} T$ with the
reservoir temperature $T$.
The force can be divided into the reversible and irreversible parts,
$\bm{f}=\bm{f}^{\rm rev}(\bm{q})+\bm{f}^{\rm ir}(\bm{q})$, according to its
behavior under time reversal as
\begin{equation}
\bm{f}^{\rm rev}(\bm{q})=\frac 1 2 (\bm{f}(\bm{q})+\bm{f}(\bm{\epsilon q})),
~\bm{f}^{\rm ir}(\bm{q})=\frac 1 2 (\bm{f}(\bm{q})-\bm{f}(\bm{\epsilon q})),
\end{equation}
where $\bm{\epsilon q}=(\bm{x},-\bm{p})$. The reversible part transforms as $\dot{\bm{p}}$
under time reversal, while the irreversible one does oppositely.
The Kramers equation for the probability density function (PDF) $\rho(\bm{q},t)$ is given by
\begin{equation}
\partial_t\rho(\bm{q},t) =
-\Big[\partial_{\bm{x}}\cdot\frac{\bm{p}}{m} +\partial_{\bm{p}} \cdot
\left(-\mathsf{G}\frac{\bm{p}}{m}+\bm{f}(\bm{q})-\mathsf{D}\partial_{\bm{p}}\right)\Big]\rho(\bm{q},t).
\label{kramers}
\end{equation}

As we shall see later, the discussion in this paper can also be applied to a more general
continuous stochastic dynamics described by the Fokker-Planck equation,
\begin{equation}
\partial_t\rho(\bm{q},t)
=\left[ -\partial_i A_i(\bm{q})+\partial_i\partial_j D_{ij}(\bm{q})\right] \rho(\bm{q},t),
\label{gen_fp}
\end{equation}
for dynamical variables $\bm{q}=(q_1,q_2,\cdots,q_N)$,
where $\partial_i\equiv\partial/\partial q_i$ and the summation convention is used throughout this paper.
The behavior under the time reversal of
$\bm{q}$ is described by the parity $\epsilon_i=+1$ or $-1$ for $q_i$.
We denote $\bm{\epsilon q}=(\epsilon_1 q_1, \epsilon_2 q_2, \cdots,\epsilon_N q_N)$.
For a general diffusion matrix $D_{ij}(\bm{q})$,
this corresponds to a set of Langevin equations with multiplicative noises.
As in the Brownian dynamics, we separate the
drift term into reversible and irreversible parts:
$A_i(\bm{q})=A_i^{\rm rev}(\bm{q})+A_i^{\rm ir}(\bm{q})$, where
\begin{align}
& A_i^{\rm rev}(\bm{q})=\frac 1 2 ( A_i(\bm{q})-\epsilon_i A_i(\bm{\epsilon q})), \\
& A_i^{\rm{ir}}(\bm{q})=\frac 1 2 (A_i(\bm{q})+\epsilon_i A_i(\bm{\epsilon q})).
\end{align}
Note that no summation convention is taken for $\epsilon_i$.
It is convenient to use the transition rate
\begin{equation}
 \omega[\bm{q},\bm{q}']=\left[ -\partial_i A_i(\bm{q'})+\partial_i\partial_j D_{ij}(\bm{q}')\right]\delta(\bm{q}-\bm{q}') ,
   \label{omega}
\end{equation}
for which Eq.~(\ref{gen_fp}) is written as $\partial_t\rho=\int d\bm{q}'\; \omega[\bm{q},\bm{q}'] \rho(\bm{q}',t)$.
The stationary state $\rho^{\rm s}(\bm{q})$ satisfies
$\int d\bm{q}'\omega[\bm{q},\bm{q}']\rho^{\rm s}(\bm{q}')=0$.

Given the forward path probability density $\mathcal{P}[\bm{q}(t)]$ for
a stochastic path $\bm{q}(t)$ for $0\le t\le\tau$,
the integral FT (IFT) holds for an arbitrary function $\mathcal{R}[\bm{q}(t)]$
of the path which has the form
\begin{equation}
 \mathcal{R}[\bm{q}(t)]=\ln
 \frac{   \mathcal{P}[\bm{q}(t)]    }
 {    \hat{\mathcal{P}}  [\hat{\bm{q}}(t)] } , \label{logratio}
\end{equation}
where $\hat{\mathcal{P}}  [\hat{\bm{q}}]$ is the path probability density for the transformed
path $\hat{\bm{q}}(t)$ with a specified time dependence. The transformation
$\bm{q}(t)\to\hat{\bm{q}}(t)$ must have the Jacobian of unity. It is straightforward then to see that
the IFT, $\langle\exp(-\mathcal{R}[\bm{q}])\rangle=1$, follows from the normalization of
$\hat{\mathcal{P}}$ \cite{speck,seifert,esposito,seifert_review}.

The IFT for the total EP, $\Delta S_{\rm tot}$, in the system and the environment is obtained by
using in Eq.~(\ref{logratio}) the time reversed path
$\bm{\epsilon q}(\tau-t)$ for $\hat{\bm{q}}(t)$ with the time reversed protocol
$\lambda(\tau-t)$ for $\hat{\mathcal{P}}$.  $\mathcal{P}$ is written as a product of the initial PDF $ \rho(\bm{q}(0),0)$ and the conditional path probability $\Pi[\bm{q}(t);\lambda(t)]$ for the system to evolve through the path $\bm{q}(t)$ starting from $\bm{q}(0)$ subject to the protocol $\lambda(t)$. Choosing the final PDF $\rho(\bm{q}(\tau),\tau)$ of the forward process as the initial PDF of the time-reversed process, we similarly write $\hat{\mathcal{P}}$ as a product of $\rho(\bm{q}(\tau),\tau)$ and $ \Pi^{\rm R}[\bm{\epsilon q}(\tau-t);\lambda(\tau-t)]$, where the latter is the conditional path probability for the corresponding time-reversed path starting from $\bm{\epsilon q}(\tau)$ subject to the corresponding time-reversed protocol $\lambda(\tau-t)$ indicated by the superscript R. Then we have
\begin{equation}
 \Delta S_{\rm tot}=\Delta S_{\rm sys}+\Delta S_{\rm env},
 \label{S_tot}
\end{equation}
where $\Delta S_{\rm sys}=-\ln \rho(\bm{q}(\tau),\tau)+\ln \rho(\bm{q}(0),0)$
is the system entropy change. The environmental EP $\Delta S_{\rm env}$ is given by
the log-ratio of the conditional probabilities as
\begin{equation}
 \Delta S_{\rm env}=\ln\frac{ \Pi[\bm{q}(t);\lambda(t)]   }
 {  \Pi^{\rm R}[\bm{\epsilon q}(\tau-t);\lambda(\tau-t)]  }.
\end{equation}
It is more convenient to consider the log-ratio of the conditional probabilities during the infinitesimal
time interval $dt$. The environmental EP for the whole time interval can be obtained by integrating
the following quantity over $[0,\tau]$:
\begin{equation}
d S_{\rm env}= \ln\frac{\Gamma[\bm{q}^\prime,t+dt|\bm{q},t]}
{\Gamma[\bm{\epsilon q},t+dt|\bm{\epsilon q}^\prime,t]},   \label{dS_env}
\end{equation}
where
\begin{equation}
\Gamma[\bm{q}^\prime,t+dt|\bm{q},t]=\delta(\bm{q}^\prime-\bm{q})+(dt)\omega[\bm{q}^\prime,\bm{q}]
\end{equation}
is the conditional probability
from the state $\bm{q}$ to $\bm{q}^\prime$
during $d t$. The conditional path probability $\Pi$ for the whole time interval is given by the infinite product of the conditional probabilities $\Gamma$'s in the limit $dt\to 0$. The protocol is implicit in $\omega[\bm{q}^\prime,\bm{q}]$. Note that the values of the protocols for the forward and the time-reversed processes are chosen to be identical in this time interval, given by $\lambda(t)$, and the superscript R is not necessary.

For the general Fokker-Planck equation, Eq.~(\ref{gen_fp}), the conditional probability
is given by the Onsager-Machlup form as \cite{lubensky,wissel}
\begin{eqnarray}
&&\Gamma [\bm{q}^\prime,t+dt|\bm{q},t]=
\frac{1}{(4\pi   dt)^{N/2}|\det(\mathsf{D}^{(\alpha)})|^{1/2}}
\label{gamma_gen}
\\
&&\times \exp\left[-\frac{dt}{4} H_i^{(\alpha)} (D^{(\alpha)})^{-1}_{ij}H_j^{(\alpha)}\right.\nonumber\\
&&\left. ~~~~~~~~~ -\alpha(dt)\partial_i A_i^{(\alpha)}+\alpha^2(dt)\partial_i\partial_j D_{ij}^{(\alpha)}\right],
 \nonumber
\end{eqnarray}
where
\begin{equation}
 H_i=\dot{q}_i-A_i+2\alpha \partial_j D_{ij}
\end{equation}
with $\bm{\dot{q}}=(\bm{q}^\prime - \bm{q})/dt$. $\alpha$ indicates the discretization scheme for which all the functions are evaluated at $\bm{q}^{(\alpha)}=\bm{q}+\alpha (\bm{q}^\prime-\bm{q})$.
$\alpha=0$ $(1/2)$ corresponds to the It\^o (Stratonovich) discretization.
For the Brownian dynamics, $D_{\bm{x}\bm{x}}=D_{\bm{x}\bm{p}}=0$, which gives
a factor of delta function enforcing the equation for $\dot{\bm{x}}$ in Eq.~(\ref{x_dot}).
Then reducing the diffusion matrix as $\mathsf{D}=\{D_{\bm{pp}}\}$,  the conditional path probability for the Brownian dynamics is given as
\begin{eqnarray}
&&\Gamma [\bm{q}^\prime,t+dt|\bm{q},t]=
\frac{\delta(\bm{x}^\prime -\bm{x} -dt (\bm{p}^{(\alpha)}/m))}
{(4\pi   dt)^{d/2}|\det(\mathsf{D}^{(\alpha)})|^{1/2}}
\label{gamma}\\
&&\times\exp\left[-\frac{dt}{4}
\bm{H}^{(\alpha)}\cdot
{\mathsf{D}^{(\alpha)}}^{-1}\cdot
\bm{H}^{(\alpha)}\right.
\nonumber\\
&&\left. ~~~~~~~~~ -\alpha dt\,\partial_{\bm{p}}\!\cdot\!\left(\! -\mathsf{G}\cdot\frac{\bm{p}^{(\alpha)}}{m}
+\bm{f}^{(\alpha)}\!\right)\right],\nonumber
\end{eqnarray}
where $\bm{H}^{(\alpha)}=\bm{\dot{p}}+\mathsf{G}\cdot\bm{p}^{(\alpha)}/m-\bm{f}^{(\alpha)}$
with $\bm{f}^{(\alpha)}=\bm{f}(\bm{q}^{(\alpha)})$.

The environmental EP can now be calculated from Eq.~(\ref{dS_env}).
For the general Fokker-Planck case with multiplicative noises,
we first note that in order to have a sensible EP for a finite time, $dS_{\rm env}$
in Eq.~(\ref{dS_env}) must be $\mathcal{O}(dt)$. In particular, $|\det(\mathsf{D}(\bm{q}))|$ must be equal to $|\det(\mathsf{D}(\bm{\epsilon q}))|$.
If there is no relation between $\mathsf{D}(\bm{q})$
and $\mathsf{D}(\bm{\epsilon q})$, there is no guarantee that
the log-ratio of the two determinants produces a $\mathcal{O}(dt)$-result.
Therefore, we need a restriction on the time reversal property of $D_{ij}(\bm{q})$.
In the following, we will restrict our discussion to the case where
\begin{equation}
\epsilon_i\epsilon_j D_{ij}(\bm{\epsilon q})=D_{ij}(\bm{q}). \label{D_cond}
\end{equation}
The same restriction and its simpler version for the diagonal $D_{ij}$ have
been used in Ref.~\cite{spinney}. As mentioned in that reference,
we also believe that all physically meaningful models are covered by this condition.
In Eq.~(\ref{dS_env}), the forward and reverse path probabilities depend on their own
discretization parameters, which we call $\alpha$ and $\beta$, respectively.
It has been shown \cite{spinney} that, if $\alpha+\beta=1$, i.e. if the same discretized points are used in both forward and reverse paths, the expression of $dS_{\rm env}$ in Eq.~(\ref{dS_env})
is independent of the discretization parameters.
The result is \cite{spinney}
\begin{eqnarray}
d S_{\rm env}&=&dt
\Big(\dot{q}_i-A_i^{\rm rev}(\bar{\bm{q}})\Big) D^{-1}_{ij}(\bar{\bm{q}})
\Big(A^{\rm ir}_j(\bar{\bm{q}}) -\partial_k D_{jk}(\bar{\bm{q}})\Big)\nonumber\\
&&-dt\;\partial_i A_i^{\rm rev}(\bar{\bm{q}}), \label{dS_env2}
\end{eqnarray}
where all the expressions turn out to be evaluate at the midpoint value $\bar{\bm{q}}=(\bm{q}+\bm{q}')/2$, independent of $\alpha$.
This particular combination of discretization parameters, $\alpha+\beta=1$, is actually due to the multiplicative noise.
For the Brownian dynamics, where the noise is additive,
$\alpha$ and $\beta$ can be arbitrary and
$dS_{\rm env}$ in Eq.~(\ref{dS_env}) is always independent of the choice of the discretization schemes \cite{kwon}.
For the Brownian dynamics, we have \cite{kylp}
\begin{eqnarray}
d S_{\rm env}&=&
dt \Big(\dot{\bm{p}}-\bm{f}^{\rm rev}(\bar{\bm{q}})\Big) \cdot\mathsf{D}^{-1}\cdot
\Big(-\mathsf{G}\frac{\bar{\bm{p}}}{m}+\bm{f}^{\rm ir}(\bar{\bm{q}})\Big)
 \nonumber\\
&&-dt\;\partial_{\bm{p}} \bm{f}^{\rm rev}(\bar{\bm{q}}). \label{dS_env1}
\end{eqnarray}
The physical meaning of the above expression has been investigated and
it was found that, when $\bm{f}(\bm{q})$ depends on the momentum,
there is an unconventional contribution to $d S_{\rm env}$
in addition to the usual heat production into the reservoir \cite{kylp}.

It is more illuminating to calculate the average of the above quantities.
The average of arbitrary quantities which are functions of $\bm{q}$ and $\bm{q}^\prime$
at time $t$ and $t+dt$, respectively, is defined by
\begin{equation}
\langle B(\bm{q}^\prime)C(\bm{q}) \rangle =\int d\bm{q}^\prime \int d\bm{q}
B(\bm{q}^\prime)\Gamma[\bm{q}^\prime,t+dt|\bm{q},t] C(\bm{q}) \rho(\bm{q}).
\label{average}
\end{equation}
In the following, the averages are
expressed in terms of the currents $\bm{j}(\bm{q})$ which are obtained from
Eq.~(\ref{gen_fp}) as
$\partial_t\rho(\bm{q},t)=-\partial_i j_i(\bm{q})$ or from Eq.~(\ref{kramers}) as
$\partial_t \rho(\bm{q},t)=-\partial_{\bm{x}}\cdot \bm{j}_{\bm{x}}-\partial_{\bm{p}}\cdot \bm{j}_{\bm{p}}$.
As for the force, we separate the currents into reversible and irreversible parts as
$\bm{j}=\bm{j}^{\rm rev}+\bm{j}^{\rm ir}$.
We have explicitly
\begin{eqnarray}
&&j_i^{\rm rev}(\bm{q})=A_i^{\rm rev}(\bm{q})\rho(\bm{q}), \label{j_rev}\\
&&j_i^{\rm ir}(\bm{q})=(A_i^{\rm ir}(\bm{q})-\partial_j D_{ij}(\bm{q}))\rho(\bm{q}). \label{j_ir}
\end{eqnarray}
For the Brownian dynamics,
$\bm{j}^{\rm rev}_{\bm{x}} =(\bm{p}/m) \rho(\bm{q})$, $\bm{j}^{\rm ir}_{\bm{x}}=\bm{0}$, and
\begin{eqnarray}
&&\bm{j}^{\rm rev}_{\bm{p}} =\bm{f}^{\rm rev}(\bm{q})\rho(\bm{q}), \label{j_rev:brown}\\
&&\bm{j}^{\rm ir}_{\bm{p}}=\left( -\mathsf{G}\frac{\bm{p}}{m}+\bm{f}^{\rm ir}(\bm{q})
-\mathsf{D}\partial_{\bm{p}} \right)\rho(\bm{q}). \label{j_ir:brown}
\end{eqnarray}
The average rate of the total EP can then be calculated from
Eqs.~(\ref{S_tot}) and (\ref{average}) as \cite{spinney,kylp}
\begin{equation}
\left\langle \frac{dS_{\rm tot}}{dt}\right\rangle=\int d\bm{q}\; \frac{j^{\rm ir}_i(\bm{q}) \;
D^{-1}_{ij}(\bm{q}) \; j^{\rm ir}_j (\bm{q})  }{\rho(\bm{q}) }. \label{totave}
\end{equation}
For the Brownian dynamics, we have a similar expression involving
only the momentum component of the irreversible
current $\bm{j}^{\rm ir}_{\bm{p}}$  as
$D^{-1}_{ij}$ exists only in that space.
The positivity of Eq.~(\ref{totave}) comes from
the positive-definiteness of $\mathsf{D}$. As we will discuss later, $\bm{j}^{\rm ir}=0$ if the DB
condition is satisfied.

When stationary states are involved, another type of EP which satisfies
the IFT can be considered.
The excess EP $\Delta S_{\rm excess}$ arising from transitions between stationary states can be constructed by
using the adjoint or dual dynamics \cite{spinney,seifert_review}.
For a given stochastic process described by $\omega$,
an adjoint process $\omega^\ast$, called $\ast$-process, can be defined as
\begin{equation}
\omega^\ast[\bm{q}^\prime,\bm{q}]\equiv\omega[\bm{q},\bm{q}^\prime]
\frac{\rho^{\rm s}(\bm{q}^\prime)}{\rho^{\rm s}(\bm{q})} , \label{gammaast}
\end{equation}
where $\rho^{\rm s}(\bm{q})$ is defined at a given $t$ as the expected stationary distribution 
if the protocol $\lambda(t)$ is kept unchanged such that 
 $[-\partial_iA_i(\bm{q},\lambda(t))+\partial_i\partial_j]\rho^{\rm s}(\bm{q})=0$.   The associated transition probability, $\Gamma^\ast[\bm{q}^\prime,t+dt|\bm{q},t]=\delta(\bm{q}'-\bm{q})+(dt) \omega^\ast[\bm{q}^\prime,\bm{q}]$, leads to
$\Gamma[\bm{q},t+dt|\bm{q}^\prime,t]\left(\rho^{\rm s}(\bm{q}^\prime)/\rho^{\rm s}(\bm{q})\right)$,
where the protocols for $\rho^{\rm s}(\bm{q})$ and $\rho^{\rm s}(\bm{q}')$ are chosen to take the same value $\lambda(t)$ at time $t$.

As in the even-variable only case,
the excess EP
in the presence of odd-parity variables can also be obtained by using the adjoint process $\Gamma^\ast$
in the following way \cite{spinney,lkp}.
For an infinitesimal time interval,
we define the excess EP as
\begin{equation}
 d S_{\rm excess}=\ln\frac{\rho(\bm{q})\Gamma[\bm{q}^\prime,t+dt|\bm{q},t]}{\rho(\bm{q}')\Gamma^\ast[\bm{q},t+dt|\bm{q}^\prime,t]}.
\label{dS_excess}
\end{equation}
We can write $d S_{\rm excess}=\ln \left(\rho(\bm{q})/\rho(\bm{q}')\right)+d S_1$ where
\begin{equation}
d S_1= \ln\frac{\Gamma[\bm{q}^\prime,t+dt|\bm{q},t]}{\Gamma^\ast[\bm{q},t+dt|\bm{q}^\prime,t]}
=\ln \frac{\rho^{\rm s}(\bm{q}^\prime)}{\rho^{\rm s}(\bm{q})},
\label{dS_1}
\end{equation}
which is the same as found in \cite{spinney,lkp}. If $\mathcal{P}^\ast$ is given from $\Gamma^*$ in the denominator of
Eq.~(\ref{logratio}), it represents a well-defined path probability since $\int d\bm{q}' \Gamma^\ast[\bm{q}^\prime,t+dt|\bm{q},t]=1$, which can be shown from the property that $\int d\bm{q}'\omega[\bm{q},\bm{q}']\rho^{\rm s}(\bm{q}')=0$. Therefore we can see that
the excess EP for a finite time interval,  $\Delta S_{\rm excess}=\int dS_{\rm excess}$ satisfies the IFT. 

Using Eq.~(\ref{dS_1}), we can readily find that
\begin{equation}
 d S_1 =-dt\;\dot{q}_i \partial_i \phi(\bar{\bm{q}}),
 \label{dS_11}
\end{equation}
where $\rho^{\rm s}(\bm{q})=\exp(-\phi(\bm{q}))$.
For the Brownian dynamics, we have
$ d S_1=-dt\big(\dot{\bm{p}}\partial_{\bm{p}}\phi(\bar{\bm{q}})
 +\dot{\bm{x}}\partial_{\bm{x}}\phi(\bar{\bm{q}})\big)$.
One can choose any distribution functions for $\rho(\bm{q})$ and $\rho(\bm{q}')$ in Eq.~(\ref{dS_excess}). If the initial and final distributions are chosen as $\rho^{\rm s}$, then the total excess EP is given as
$ \Delta S_{\rm excess}=\int^\tau_0 dt\;  \dot{\lambda}(t)\left(\partial\phi/\partial \lambda\right)$,
which is exactly the familiar Hatano-Sasa expression \cite{hatano}.

If $\rho(\bm{q}')$ is chosen as the PDF at time $t+dt$ given the PDF $\rho(\bm{q})$ at time $t$,
then we have $dS_{\rm excess}=dS_{\rm sys}+dS_1$. In this case, the total EP is rearranged as
$dS_{\rm tot}=d S_{\rm excess}+d S_{\rm hk}$ from Eq.~(\ref{S_tot}), where
the remaining $d S_{\rm hk}$, so-called the house-keeping EP, will be introduced in the next section.
The average excess EP rate can also be obtained as for the environmental EP. First, we note
$\langle dS_{\rm sys}/dt \rangle=-\int d\bm{q}(\partial_t\rho(\bm{q}))\ln \rho(\bm{q})$.
Using Eqs.~(\ref{average}) and (\ref{dS_11}),
we get $\langle dS_1/dt\rangle=\int d\bm{q} j_i(\bm{q})\partial_i\ln\rho^{\rm s}(\bm{q})
=\int d\bm{q}(\partial_t\rho(\bm{q}))\ln \rho^{\rm s}(\bm{q})$. Then, one can show \cite{spinney} that
\begin{eqnarray}
  &&\left\langle\frac{d  S_{\rm excess}}{dt}\right \rangle
  =-\int d\bm{q}\; (\partial_t \rho(\bm{q}))\ln\frac{\rho(\bm{q})}{\rho^{\rm s}(\bm{q})}
\label{S_ex:ave} \\
  &&=
  \int d\bm{q}\; \Big\{ \frac{j_i(\bm{q})}{\rho(\bm{q})}
  -\frac{j^{\rm s}_i(\bm{q})}{\rho^{\rm s}(\bm{q})} \Big\}
  D^{-1}_{ij}(\bm{q}) \Big\{ \frac{j_j(\bm{q})}{\rho(\bm{q})}
  -\frac{j^{\rm s}_j(\bm{q})}{\rho^{\rm s}(\bm{q})}\Big\}
  \rho(\bm{q}), \nonumber
\end{eqnarray}
where the stationary state currents $\bm{j}^{\rm s}=\bm{j}^{\rm s,rev}+\bm{j}^{\rm s,ir}$ are defined similarly
to Eqs.~(\ref{j_rev}) and (\ref{j_ir}).
One just replaces $\rho(\bm{q})$ by $\rho^{\rm s}(\bm{q})$ in the those expressions.
For the Brownian dynamics, the above expression again involves only the momentum component of the currents.
The first line of Eq.~(\ref{S_ex:ave}) explicitly exhibits a transient nature of the excess EP rate.

\section{Housekeeping Entropy Production}
\label{sec:hk}

We now discuss the main subject of the present paper.
The remaining part of the total entropy apart from the excess EP is the housekeeping EP:
$\Delta S_{\rm hk}=\Delta S_{\rm tot}-\Delta S_{\rm excess}$.
In the absence of odd-parity variables, the housekeeping EP satisfies the IFT. 
From Eqs.~(\ref{dS_env}) and (\ref{dS_1}), however, we can show that the housekeeping entropy cannot be written
as the ratio of two path probabilities, and therefore does not satisfy the IFT. This is in contrast to 
the case where all the variables have even parity.

Although the whole housekeeping EP does not
satisfy the IFT, one can identify a part of housekeeping EP that satisfies the IFT.
In a discrete-jumping process, this EP denoted by $\Delta S_{\rm bDB}$ was calculated and
shown to be directly responsible for the breakage of the DB in the stationary state \cite{lkp}.
For the continuous stochastic dynamics, a part of the housekeeping EP denoted by $\Delta S_2$ was separated out and
shown to satisfy the IFT \cite{spinney}. It was obtained by
adding up the infinitesimal contributions,
\begin{equation}
 dS_2= \ln\frac{\Gamma[\bm{q}^\prime,t+dt|\bm{q},t]}{\Gamma^\ast[\bm{\epsilon q}^\prime,t+dt|\bm{\epsilon q},t]}.
 \label{dS_2}
\end{equation}
It is not obvious, however, how the expression in Eq.~(\ref{dS_2}) is related to the broken DB
as we shall see in more detail below.
In the following, we will show that a direct attempt to construct
$\Delta S_{\rm bDB}$ for the case of continuous stochastic dynamics
poses serious problems. In the subsequent subsections,
we instead find a series of EPs each of which is a part of housekeeping EP and satisfies the IFT.
$\Delta S_2$ found in Ref.~\cite{spinney} is one of them. We will show that all these EPs
are associated with the DB breakage in the stationary state under a very general assumption.

\subsection{Problems with $\Delta S_{\rm bDB}$ for continuous stochastic dynamics}

In the presence of odd-parity variables, the DB condition reads
\begin{equation}
\omega[\bm{q}^\prime,\bm{q}]\rho^{\rm s}(\bm{q})=
\omega[\bm{\epsilon q},\bm{\epsilon q}^\prime]\rho^{\rm s}(\bm{\epsilon q}^\prime).
\label{db}
\end{equation}
In order to measure the departure from the DB, we define the adjoint $\dag$-process
\begin{equation}
 \omega^\dag[\bm{q}',\bm{q}]\equiv\omega[\bm{\epsilon q},\bm{\epsilon q}^\prime]\frac{\rho^{\rm s}(\bm{\epsilon q}^\prime)}
 {\rho^{\rm s}(\bm{q})}
 \label{omega_dag}
\end{equation}
such that the DB condition is equivalent to the condition $\omega[\bm{q}',\bm{q}]=\omega^\dag[\bm{q}',\bm{q}]$.
This is a well-defined stochastic process as one can easily see that
$ \int d\bm{q}' \omega^\dag[\bm{q}',\bm{q}]=0$
follows from the stationarity of $\rho^{\rm s}$.
In terms of the transition probability, we have
\begin{eqnarray}
\Gamma^\dagger[\bm{q}^\prime,t+dt|\bm{q},t]&\equiv&\Gamma[\bm{\epsilon q},t+dt|\bm{\epsilon q}^\prime ,t]
\frac{\rho^{\rm s} (\bm{\epsilon q}^\prime)}
{\rho^{\rm s}(\bm{q})}\nonumber \\
&&+\delta(\bm{q}-\bm{q}^\prime)\left(
1-\frac{\rho^{\rm s}(\bm{\epsilon q})}
{\rho^{\rm s}(\bm{q})}\right). \label{daggerprocess}
\end{eqnarray}
One may expect that
the EP associated with the breakage of DB can then be measured by considering the path probability ratio
\begin{equation}
d S_{\rm bDB}\equiv \ln\frac{\Gamma[\bm{q}^\prime,t+dt|\bm{q},t]}
{\Gamma^\dagger[\bm{q}^\prime,t+dt|\bm{q},t]}.
\label{SbDB}
\end{equation}
For a discrete jumping process in the presence of odd-parity variables \cite{lkp},
this quantity has been calculated and
the corresponding EP $\Delta S_{\rm bDB}$ during a finite time interval has been shown to
satisfy the IFT.
However, this procedure cannot be repeated for a continuous stochastic dynamics.

We first look at the $\dagger$-process more closely.
Using the quantum mechanical description, we write a term in the right hand side of Eq.~(\ref{omega_dag}) as
\begin{align}
& \omega\big[\bm{\epsilon q},\bm{\epsilon q}^\prime\big]\rho^{\rm s}(\bm{\epsilon q}^\prime)   \\
&=\langle \bm{q}| \left[\{ -\epsilon_i \hat{r}_i A_i(\bm{\epsilon \hat{q}})+\epsilon_i\epsilon_j\hat{r}_i\hat{r}_j
 D_{ij}(\bm{\epsilon \hat{q}})\}\rho^{\rm s}(\bm{\epsilon \hat{q}})\right]|\bm{q}'\rangle. \nonumber
\end{align}
where $\hat{q}_i$,  $\hat{r}_i$ are non-commuting operators, satisfying $[\hat{r}_j,\hat{q}_i]=\delta_{ij}$, and have the properties: $\langle\bm{q}|\hat{r}_i|\bm{q}'\rangle=\partial_i\delta(\bm{q}-\bm{q}')$ and $\langle\bm{q}|\hat{q}_i|\bm{q}'\rangle=q_i\delta(\bm{q}-\bm{q}')$.
We can then move $\hat{r}_i$'s to the right inside the bracket using the commutator relation. As a result we have three terms proportional to $\delta(\bm{q}-\bm{q}')$, $\partial_i \delta(\bm{q}-\bm{q}')$, and $\partial_i\partial_j \delta(\bm{q}-\bm{q}')$, respectively. Note that the term proportional to $\delta(\bm{q}-\bm{q}')$ vanishes because of the stationary condition of $\rho^{\rm s}$.
From Eq.~(\ref{omega_dag}), we have
\begin{align}
 \omega^\dag[\bm{q}',\bm{q}]=&\Big[ -\epsilon_i A_i (\bm{\epsilon q})e^{\phi_{\rm A}(\bm{q})}
 \nonumber \\
+ & \frac{2}{\rho_{\rm s}(\bm{q})}\epsilon_i\epsilon_j \big(\partial_j D_{ij}(\bm{\epsilon q})\rho_{\rm s}(\bm{\epsilon q})\big)
 \Big]\partial_i \delta(\bm{q}-\bm{q}')
 \nonumber \\
 &+\big[\epsilon_i\epsilon_j D_{ij}(\bm{\epsilon q})e^{\phi_{\rm A}(\bm{q})}\big]
 \partial_i\partial_j \delta(\bm{q}-\bm{q}'),
\label{omega_calc}
\end{align}
where
\begin{equation}
 \phi_{\rm A}(\bm{q})= \phi(\bm{q})-\phi(\bm{\epsilon q})=\ln\frac{\rho^{\rm s}(\bm{\epsilon q})}{\rho^{\rm s}(\bm{ q})}.
\end{equation}
This dagger process can be put into the standard form like Eq.~(\ref{omega}) by noting that
\begin{align}
\omega^\dag[\bm{q}',\bm{q}]=&\left[ -\partial'_i A^\dag_i(\bm{q})
  +\partial'_i\partial'_j D^\dag_{ij}(\bm{q})\right]\delta(\bm{q}'-\bm{q})
\nonumber \\
  =&   \left[ A^\dag_i(\bm{q}) \partial_i
  + D^\dag_{ij}(\bm{q})\partial_i\partial_j\right]\delta(\bm{q}'-\bm{q}).
\end{align}
By comparing the two expressions for $\omega^\dag$, we have
\begin{align}
 &A_i^\dag (\bm{q})= -\epsilon_i A_i (\bm{\epsilon q})e^{\phi_{\rm A}(\bm{q})}
+ \frac{2\epsilon_i\epsilon_j }{\rho^{\rm s}(\bm{q})}\partial_j D_{ij}(\bm{\epsilon q})\rho^{\rm s}(\bm{\epsilon q}),
\label{A_dag}\\
& D^\dag_{ij}(\bm{q})=\epsilon_i\epsilon_j D_{ij}(\bm{\epsilon q})e^{\phi_{\rm A}(\bm{q})}.
\label{D_dag}
\end{align}
Using the restriction on $\mathsf{D}$, Eq.~(\ref{D_cond}), we have
\begin{align}
 &A_i^\dag (\bm{q})= -\epsilon_i A_i (\bm{\epsilon q})e^{\phi_{\rm A}(\bm{q})}
+ \frac{2 }{\rho^{\rm s}(\bm{q})}\partial_j D_{ij}(\bm{q})\rho^{\rm s}(\bm{\epsilon q}),
\label{A_dag_1}\\
& D^\dag_{ij}(\bm{q})= D_{ij}(\bm{q})e^{\phi_{\rm A}(\bm{q})}.
\label{D_dag_1}
\end{align}

Explicit expressions for the short-time transition probability
$\Gamma^\dag[\bm{q}^\prime,t+dt|\bm{q},t]$
can be obtained  in Eq.~(\ref{gamma_gen}) by
using $A^\dagger$ and $D^\dagger$ in places of $A$ and $D$, respectively.
Now when we try to calculate $d S_{\rm bDB}$ directly from Eq.~(\ref{SbDB}), we note that
the presence of $e^{\phi_{\rm A}}$ factor in Eq.~(\ref{D_dag_1}) makes the two processes $\Gamma$
and $\Gamma^\dag$ with different multiplicative noises.  In particular,
$\ln |\det (e^{\phi_{\rm A}(\bar{\bm{q}} )} \mathsf{D}(\bar{\bm{q}})  )/\det (\mathsf{D}(\bar{\bm{q}}))|^{1/2}$ resulting from Eq.~(\ref{SbDB}) is clearly not of $\mathcal{O}(dt)$.
This in turn makes $\Delta S_{\rm bDB}$ for a finite time interval diverge and become ill-defined.

For the Brownian dynamics, we have from Eqs.~(\ref{gamma}), (\ref{A_dag_1}),
and (\ref{D_dag_1})
\begin{eqnarray}
&&\Gamma^\dagger [\bm{q}^\prime,t+dt|\bm{q},t]=
\frac{\delta(\bm{x}^\prime -\bm{x} -dt (\bm{p}^{(\alpha)}/m)e^{\phi_{\rm A}^{(\alpha)}})}
{(4\pi  e^{\phi_{\rm A}} dt)^{d/2}|\det(\mathsf{D}^{(\alpha)})|^{1/2}}
\nonumber \\
&&\times \exp\Bigg[-\frac{dt}{4e^{\phi_{\rm A}^{(\alpha)}}}
\bm{H}^{\dagger(\alpha)}\cdot
{\mathsf{D}^{(\alpha)}}^{-1}\cdot\bm{H}^{\dagger(\alpha)}
\nonumber\\
&&-\alpha dt \partial_{\bm{x}}\cdot\Big(e^{\phi_{\rm A}^{(\alpha)}} \frac{\bm{p}^{(\alpha)}}{m} \Big) -\alpha dt\, \partial_{\bm{p}}\cdot\Big\{e^{\phi_{\rm A}^{(\alpha)}}\Big( \mathsf{G}\cdot\frac{\bm{p}^{(\alpha)}}{m}
\nonumber\\
&&
 +\bm{f}(\bm{\epsilon} \bm{q}^{(\alpha)})
 -2\mathsf{D}\cdot\partial_{\bm{p}}\phi(\bm{\epsilon} \bm{q}^{(\alpha)})\Big)\Big\}
 \nonumber\\
&&\quad
+\alpha^2 dt\, \partial_{\bm{p}}\cdot\mathsf{D}\cdot\partial_{\bm{p}}e^{\phi_{\rm A}^{(\alpha)}}\Bigg],\label{gammadagger}
\end{eqnarray}
where
\begin{eqnarray}
\bm{H}^{\dagger}&\equiv&\bm{\dot{p}} -e^{\phi_{\rm A}}
 \Big( \mathsf{G}\cdot\frac{\bm{p}}{m} +\bm{f}( \bm{\epsilon q}) \nonumber \\
&& \quad\quad -2\mathsf{D}\cdot
\partial_{\bm{p}}\phi(\bm{\epsilon q})-2\alpha \mathsf{D}\cdot\partial_{\bm{p}}\phi_{\rm A}(\bm{q})\Big).
\end{eqnarray}
Here even if we start from an additive noise in the
forward process, the dagger process becomes one with a multiplicative noise.
Again, we have the same problem with the determinants. In addition,
we have a delta function enforcing the equation of motion,
\begin{equation}
 \bm{\dot{x}}=\frac{\bm{p}}{m} e^{\phi_{\rm A}(\bm{q})}
\end{equation}
for the $\dag$-process.
This signifies the mismatch between the forward path and the one described by the $\dagger$-process.

From the above discussion, it seems impossible to construct the EP for the continuous dynamics associated with the breakage of the DB directly from the discrete stochastic dynamics.
Below we introduce another stochastic process for which the path probability ratio
taken with the forward process for a finite time interval is well defined.
We also want this EP to represent the departure from the DB. Before we proceed,  we need to investigate the DB condition in the presence of odd-parity variables more carefully.

\subsection{The DB condition}

As mentioned above, the DB condition
is equivalent to setting $\omega=\omega^\dagger$. If $\rho^{\rm s}$
satisfies the stationary distribution, this condition reduces to
$A_i^\dag=A_i$ and $D^\dag_{ij}=D_{ij}$
in Eqs.~(\ref{A_dag}) and (\ref{D_dag}). We note that
in a discrete jumping process, the DB and the parity symmetry of the stationary distribution are
two independent  conditions \cite{lkp}.
From the second condition Eq.~(\ref{D_dag}), we can see that
the parity symmetry, $\phi_{\rm A}(\bm{q})=0$, is in general not guaranteed unless
the diffusion matrix satisfies Eq.~(\ref{D_cond}).
If this relation holds, as we are assuming in this paper, then the DB is equivalent to
the parity symmetry. The vanishing irreversible current in the stationary state,
\begin{equation}
 j^{\rm s,ir}_i(\bm{q})\equiv A_i^{\rm ir}(\bm{q})\rho^{\rm s}(\bm{q})
 -\partial_jD_{ij}(\bm{q})\rho^{\rm s}(\bm{q})=0 \label{jir=0}
\end{equation}
also follows from the first condition Eq.~(\ref{A_dag_1}).
We note that, in conventional
textbooks \cite{risken}, the DB condition is defined as
$\omega=\omega^\dagger$ {\em and} $\rho^{\rm s}(\bm{q})=\rho^{\rm s}(\bm{\epsilon q})$.
The condition Eq.~(\ref{D_cond})
is then regarded as a requirement for the existence of DB.

Therefore, the DB condition is equivalent to the parity symmetry and Eq.~(\ref{jir=0}).
In the following, we will show that under a very broad assumption,
these two conditions are not actually independent.
Let us assume only Eq.~(\ref{jir=0}) for all $i$. Then
  \begin{equation}
   A_i^{\rm ir}(\bm{q})=\frac{1}{\rho^{\rm s}(\bm{q})}\partial_j D_{ij}(\bm{q})\rho^{\rm s}(\bm{q})
  \end{equation}
Multiplying by $\epsilon_i$ and changing $\bm{q}\to \bm{\epsilon q}$, we have
\begin{equation}
 \epsilon_i A_i^{\rm ir}(\bm{\epsilon q})=\frac{\epsilon_i \epsilon_j}{\rho^{\rm s}(\bm{\epsilon q})}
 \partial_j D_{ij}(\bm{\epsilon q})\rho^{\rm s}(\bm{\epsilon q})
\end{equation}
But using $ \epsilon_i A_i^{\rm ir}(\bm{\epsilon q})= A_i^{\rm ir}(\bm{q})$
and Eq.~(\ref{D_cond}), we have
$ D_{ij}(\bm{q})\partial_j \phi(\bm{q})= D_{ij}(\bm{q})\partial_j \phi(\bm{\epsilon q})$.
Thus Eq.~(\ref{jir=0}) implies
\begin{equation}
 D_{ij}(\bm{q})\partial_j \phi_{\rm A}(\bm{q})= 0.
 \label{del_phi_a}
\end{equation}

Now if $\mathsf{D}^{-1}$ exists, we have for all $j$, $\partial_j \phi_{\rm A} (\bm{q})=0$, or
$\phi_{\rm A}(\bm{q})=\phi_0$, a constant.
But from the normalization
   \begin{align}
    1&=\int d\bm{q}\; \rho^{\rm s}(\bm{q})=e^{-\phi_0}\int d\bm{q}\;  \rho^{\rm s}(\bm{\epsilon q}) \nonumber \\
    &=e^{-\phi_0} \int d\bm{q}^\prime\; \rho^{\rm s}(\bm{q}^\prime)=e^{-\phi_0} ,
   \end{align}
where we have used the integration variable change $\bm{q}^\prime=\bm{\epsilon q}$ and
the fact that $d\bm{q}^\prime=d\bm{ q}$. We therefore conclude that $\phi_0=0$ and that the
two conditions are {\em not} independent, but the vanishing irreversible current in stationary state implies
the parity symmetry.

For Brownian dynamics, $\mathsf{D}^{-1}$ in the whole space does not exist
since the $\bm{x}$-components of the
diffusion matrix vanish as
   \[
   D_{\bm{x}\bm{x}}=D_{\bm{x}\bm{p}}=D_{\bm{p}\bm{x}}=0,~~~D_{\bm{p}\bm{p}}=D.
   \]
Therefore from Eq.~(\ref{del_phi_a}) we only have $\partial_{\bm{p}}\phi_{\rm A}(\bm{x},\bm{p})=0$ or
   \begin{equation}
   \phi(\bm{x},\bm{p})-\phi(\bm{x},-\bm{p})=\phi_0(\bm{x})
   \end{equation}
Following a similar discussion to above, we have
   \begin{align}
   &\int d\bm{p}\; \rho^{\rm s}(\bm{x},\bm{p})=e^{-\phi_0(\bm{x})}\int d\bm{p}\; \rho^{\rm s}(\bm{x},-\bm{p}) \nonumber \\
   &=e^{-\phi_0(\bm{x})}\int d\bm{p}\; \rho^{\rm s}(\bm{x},\bm{p}).
   \end{align}
We again conclude that $\phi_0(\bm{x})=0$ and the parity symmetry of $\rho^{\rm s}$.
We can easily generalize this discussion
to the case of several variables where $\mathsf{D}^{-1}$ exists at least in the subspace spanned
by the odd-parity variables. We believe this is a reasonable assumption for a physical system
described by a combination of even and odd parity variables. This is certainly true for
the Brownian dynamics. In the following discussion,
we will assume this to be true.

\subsection{Generalized adjoint process}

We now introduce a generalization of the $\dagger$-process for which the path probability ratio
taken with the forward process for a finite time interval is well defined. It is based
on the similar construction given in Ref.~\cite{gen_adj}.  For
given $\omega$ and an arbitrary $h(\bm{q})$, we define
\begin{align}
 &\omega_h^\ddag[\bm{q}',\bm{q}]=
 \omega[\bm{\epsilon q},\bm{\epsilon q}']\frac{h(\bm{q}')}{h(\bm{q})}  \nonumber \\
&- \delta(\bm{q}-\bm{q}')\frac{1}{h(\bm{q})}
\int d\bm{q}'' \omega[\bm{\epsilon q},\bm{\epsilon q}'']h(\bm{q}'').
 \label{omega_ddag}
\end{align}
 The last term ensures 
the stochasticity of $\omega_h^\ddag$, i.e. $\int d\bm{q}' \omega_h^\ddag[\bm{q}',\bm{q}]=0$.
If we choose $h(\bm{q})=h_0(\bm{q})\equiv\rho^{\rm s}(\bm{\epsilon q})$, then the last term
of  Eq.~(\ref{omega_ddag}) vanishes automatically and we have
$\omega_{h_0}^\ddag[\bm{q}^\prime,\bm{q}]=\omega^\ast[\bm{\epsilon q}^\prime,\bm{\epsilon q}]$
where the $\ast$-process is defined in Eq.~(\ref{gammaast}).  This was the choice
made by Spinney and Ford \cite{spinney} to construct a component of housekeeping EP, $\Delta S_2$, that satisfies the IFT.

Following the same procedure as in
Eq.~(\ref{omega_calc}), we can rewrite Eq.~(\ref{omega_ddag}) as
\begin{align}
&\omega_h^\ddag[\bm{q}',\bm{q}]=
\nonumber\\
&\big[ -\epsilon_i  A_i(\bm{\epsilon q}) +\frac{2}{h(\bm{q})}
 (\partial_j D_{ij}(\bm{q})h(\bm{q})) \big]\partial_i \delta(\bm{q}-\bm{q}') \nonumber \\
&+ D_{ij}(\bm{q})
 \partial_i\partial_j \delta(\bm{q}-\bm{q}').
\end{align}
Note that the term proportional to $\delta(\bm{q}-\bm{q}')$
is cancelled by the second term on the right hand side of
Eq.~(\ref{omega_ddag}).  Writing
$ \omega_h^\ddag[\bm{q}',\bm{q}]= [A^\ddag_i(\bm{q}) \partial_i
  + D^\ddag_{ij}(\bm{q})\partial_i\partial_j]\delta(\bm{q}'-\bm{q})$, we can identify
\begin{align}
 &A_i^\ddag (\bm{q})= -\epsilon_i A_i (\bm{\epsilon q})
+ \frac{2 }{h(\bm{q})}\partial_j (D_{ij}(\bm{q})h(\bm{q})),
\label{A_ddag}\\
& D^\ddag_{ij}(\bm{q})= D_{ij}(\bm{q}).
\label{Dij_ddag}
\end{align}
Because of Eq.~(\ref{Dij_ddag}),
$\ln\Gamma /\Gamma^\ddag$ will now be $O(dt)$.

We now study what the condition that $\omega=\omega_h^\ddag$ means. The path probability ratio
\begin{equation}
 d S_h\equiv \ln\frac{\Gamma[\mathbf{q}^\prime,t+dt|\mathbf{q},t]}
{\Gamma_h^\ddag[\mathbf{q}^\prime,t+dt|\mathbf{q},t]}
\label{dS_h}
\end{equation}
will then measure the breakage of this condition.
From Eq.~(\ref{A_ddag}), the condition $\omega=\omega_h^\ddag$
amounts to
\begin{equation}
A^{\rm ir}_i(\bm{q})-\frac{1}{h(\bm{q})}\partial_j D_{ij}(\bm{q})h(\bm{q})=0.
 \label{condition}
\end{equation}
We write this condition in terms of $ j^{\rm s,ir}_i (\bm{q})$.
From the definition of $ j^{\rm s,ir}_i (\bm{q})$ in Eq.~(\ref{jir=0}), we have the identities
\begin{align}
A^{\rm ir}_i(\bm{q})=&
 \frac{ j^{\rm s,ir}_i (\bm{q})}{\rho^{\rm s}(\bm{q})}
 +\frac{1}{\rho^{\rm s}(\bm{q})}\partial_j D_{ij}(\bm{q})\rho^{\rm s}(\bm{q}) \label{Ai_ir_1}\\
 =&\frac{ \epsilon_i j^{\rm s,ir}_i (\bm{\epsilon q})}{\rho^{\rm s}(\bm{\epsilon q})}
 +\frac{1}{\rho^{\rm s}(\bm{\epsilon q})}\partial_j D_{ij}(\bm{q})\rho^{\rm s}(\bm{\epsilon q}),
 \label{Ai_ir_2}
\end{align}
where in the second equality we have multiplied by $\epsilon_i$,
changed $\bm{q}\to \bm{\epsilon q}$, and used
$ \epsilon_i A_i^{\rm ir}(\bm{\epsilon q})= A_i^{\rm ir}(\bm{q})$
and Eq.~(\ref{D_cond}).
If we insert this expression into Eq.~(\ref{condition}),
we can rewrite the condition $\omega_h^\ddag=\omega$  as
\begin{equation}
 j^{\rm s,ir}_i (\bm{q})=\rho^{\rm s}(\bm{q})D_{ij}(\bm{q})\left[\frac{\partial_j h(\bm{q})}{h(\bm{q})}
 - \frac{\partial_j \rho^{\rm s}(\bm{q})}{\rho^{\rm s}(\bm{q})} \right] ,
 \label{cond1}
\end{equation}
or
\begin{equation}
\epsilon_i j^{\rm s,ir}_i (\bm{\epsilon q})=\rho^{\rm s}(\bm{\epsilon q})D_{ij}(\bm{q})
 \left[\frac{\partial_j h(\bm{q})}{h(\bm{q})}
 - \frac{\partial_j \rho^{\rm s}(\bm{\epsilon q})}{\rho^{\rm s}(\bm{\epsilon q})} \right] .
\label{cond2}
 \end{equation}
We can easily see for the choices of
 $h(\bm{q})=h_1(\bm{q})\equiv \rho^{\rm s}(\bm{q})$
and the one by Spinney and Ford,
$ h(\bm{q})=h_0(\bm{q})=\rho^{\rm s}(\bm{\epsilon q})$
that the condition is equivalent to $ j^{\rm s,ir}_i (\bm{q})=0$.
So in these cases, we can say that
$\Delta S_h$ measures the departure from the state of the vanishing irreversible current
in the stationary state.  From the discussion in the previous subsection,
the vanishing irreversible current in the stationary state
actually means the parity symmetry $\phi_{\rm A}=0$ and consequently the DB.
In Ref.~\cite{spinney}, $\Delta S_{h_0}$ is denoted by $\Delta S_2$.

The parity symmetry $\phi_{\rm A}=0$ also follows directly from the condition Eq.~(\ref{condition}), since
we can rewrite that equation as
\begin{equation}
 A^{\rm ir}_i(\bm{q})-\frac{1}{h(\bm{\epsilon q})}\partial_j D_{ij}(\bm{q})h(\bm{\epsilon q})=0,
 \label{condition1}
\end{equation}
by using the same manipulations as in the steps from Eq.~(\ref{Ai_ir_1}) to (\ref{Ai_ir_2}).
Therefore, for the condition $\omega=\omega_h^\ddag$ to hold, $h$
must satisfy
\begin{equation}
 D_{ij}(\bm{q})\left[\frac{\partial_j h(\bm{q})}{h(\bm{q})}-
 \frac{\partial_j h(\bm{\epsilon q})}{h(\bm{\epsilon q})}\right] =0
 \label{h_cond}
\end{equation}
For the choices of $h_0$ and $h_1$, this is just Eq.~(\ref{del_phi_a}).

We can construct more general $h(\bm{q})$ such that the condition
 $\omega=\omega_h^\ddag$ implies the DB.
Let us consider for some number $\sigma$
\begin{equation}
 h(\bm{q})=h_\sigma(\bm{q})\equiv \left( \rho^{\rm s} (\bm{q})\right)^\sigma
 \left( \rho^{\rm s} (\bm{\epsilon q})\right)^{1-\sigma} .
 \label{h_sigma}
\end{equation}
Then Eq.~(\ref{h_cond}) is rewritten as
\begin{equation}
 (1-2\sigma)  D_{ij}(\bm{q})\partial_j \phi_{\rm A}(\bm{q})= 0.
\end{equation}
For all values of $\sigma$ except for $\sigma= 1/2$, this condition again gives the parity symmetry, $\phi_{\rm A}=0$.
Using Eqs. (\ref{cond1}) and (\ref{cond2}), we obtain the condition
$\omega=\omega_{h_\sigma}^\ddag$ in terms of the irreversible currents as
\begin{equation}
 j^{\rm s,ir}_i (\bm{q})=(1-\sigma)\rho^{\rm s}(\bm{q})D_{ij}(\bm{q})\partial_j\phi_{\rm A}(\bm{q})=0,
 \label{cond1_sigma}
\end{equation}
or
\begin{equation}
 \epsilon_i j^{\rm s,ir}_i (\bm{\epsilon q})=-\sigma\rho^{\rm s}(\bm{\epsilon q})D_{ij}(\bm{q})
 \partial_j\phi_{\rm A}(\bm{q})=0,
\end{equation}
where we have assumed $\sigma\neq 1/2$ for the last equalities in both equations to hold.

\subsection{Entropy productions}

As we have seen in the previous subsection, when $h(\bm{q})=h_\sigma(\bm{q})$
for $\sigma\neq 1/2$,
the corresponding EP $d S^\sigma_{\rm{bDB}}\equiv dS_{h_\sigma}$ defined in Eq.~(\ref{dS_h})
can be regarded
as a measure of the DB breakage.
As this quantity is given by the ratio of two path probabilities,
it satisfies the IFT for all $\sigma$.
In order to calculate this EP,
We first evaluate
$\Gamma^\ddag_\sigma\equiv\Gamma^\ddag_{h_\sigma}$ by replacing $A_i(\bm{q})$
in Eq.~(\ref{gamma_gen}) by $A_i^\ddag(\bm{q})$ in Eq.~(\ref{A_ddag})
with $h(\bm{q})=h_\sigma(\bm{q})$ given in Eq.~(\ref{h_sigma}).  As before, both $\Gamma$
and $\Gamma^\ddag_\sigma$ depend on their own discretization schemes.
For general multiplicative noises,
we find that, if we use the same discretization parameter
for both processes, the resulting EP, $dS_\sigma$ is independent of this parameter.
Since both processes are forward in time direction in this case, it actually corresponds to taking the same discretized
points as in the calculation of $dS_{\rm env}$.
When the noise is additive, things are simpler.
We are free to choose the discretization parameters for both processes
as the resulting EP is always independent of the parameter anyway.
After some algebra, we obtain
the EP as
\begin{widetext}
\begin{align}
dS^\sigma_{\rm{bDB}} &=
dt\; \Big(\dot{q}_i-A_i^{\rm rev}(\bar{\bm{q}})+
D_{ik}(\bar{\bm{q}})(\partial_k \psi_\sigma(\bar{\bm{q}}))\Big) D^{-1}_{ij}(\bar{\bm{q}})
\Big(A_j^{\rm ir}(\bar{\bm{q}})-\partial_l D_{jl}(\bar{\bm{q}})+
D_{jl}(\bar{\bm{q}})\partial_l \psi_\sigma(\bar{\bm{q}})\Big)
 \nonumber\\
&-dt\;\partial_i \Big( A_i^{\rm ir}(\bar{\bm{q}})-\partial_k D_{ik}(\bar{\bm{q}})
+D_{ik}(\bar{\bm{q}})\partial_k \psi_\sigma(\bar{\bm{q}})\Big),
\label{dS_sigma}
\end{align}
where we define
\begin{equation}
\psi_\sigma(\bm{q})\equiv -\ln h_\sigma(\bm{q})
=\phi(\bm{\epsilon q})+\sigma\phi_{\rm A}(\bm{q}).
\end{equation}
For the Brownian dynamics, we obtain
\begin{eqnarray}
d S^\sigma_{\rm{bDB}}
&=&dt\;  \Big(\dot{\bm{p}}-\bm{f}^{\rm rev}(\bar{\bm{q}})
+\partial_{\bm{p}}\psi_\sigma(\bar{\bm{q}})\cdot\mathsf{D}\Big)
\cdot\mathsf{D}^{-1}\cdot\Big(-\mathsf{G}\cdot\frac{\bar{\bm{p}}}{m}+\bm{f}^{\rm ir}(\bar{\bm{q}})
+\mathsf{D}\cdot\partial_{\bm{p}} \psi_\sigma (\bar{\bm{q}}) \Big)      \nonumber \\
&&-dt\;\partial_{\bm{p}}\cdot\Big(-\mathsf{G}\cdot\frac{\bar{\bm{p}}}{m}+\bm{f}^{\rm ir}(\bar{\bm{q}})
+\mathsf{D}\cdot\partial_{\bm{p}} \psi_\sigma ( \bar{\bm{q}}) \Big). \label{dS_sigma_B}
\end{eqnarray}
Note that we again express the EP using the midpoint value $\bar{\bm{q}}$. When
$\sigma=0$, one can check that Eq.~(\ref{dS_sigma})
reduces to the EP studied in Ref.~\cite{spinney}. (Equation (59) in that paper is
the same expression as ours when $\sigma=0$ and $D_{ij}(\bm{q})=\delta_{ij}D_i(\bm{q})$.)
\end{widetext}

The average rate of this EP can be obtained in a similar way to the case of the total EP. We obtain
the following two equivalent expressions:
\begin{align}
&\left\langle \frac{d S^\sigma_{\rm{bDB}}}{dt}  \right\rangle =
\int d\mathbf{q}\;
\Big\{  \frac{\epsilon_i j_i^{\rm s,ir}(\bm{\epsilon q})}{\rho^{\rm s}(\bm{\epsilon q})}
+\sigma D_{ik}(\bm{q})
 \partial_k\phi_{\rm A}(\bm{q})
\Big\} \nonumber \\
&~~~\times D^{-1}_{ij} (\bm{ q}) \Big\{
 \frac{\epsilon_j j_j^{\rm s,ir} (\bm{\epsilon q})}{\rho^{\rm s}(\bm{\epsilon q})}
+\sigma D_{jl}(\bm{q})
 \partial_l\phi_{\rm A}(\bm{q})
\Big\}\rho(\mathbf{q}),
\label{dS_sigma_rate}
\end{align}
or
\begin{align}
&\left\langle \frac{d S^\sigma_{\rm{bDB}}}{dt}  \right\rangle =
\int d\mathbf{q}\;
\Big\{  \frac{j_i^{\rm s,ir} (\bm{q})}{\rho^{\rm s}(\bm{ q})}
-(1-\sigma)D_{ik}(\bm{q})
 \partial_k\phi_{\rm A}(\bm{q})
\Big\} \nonumber \\
&\times D^{-1}_{ij} (\bm{q}) \Big\{
 \frac{j_j^{\rm s,ir} (\bm{ q})}{\rho^{\rm s}(\bm{ q})}
 -(1-\sigma)D_{jl}(\bm{q})
 \partial_l\phi_{\rm A}(\bm{q})
\Big\}\rho(\mathbf{q}).
\label{dS_sigma_rate_1}
\end{align}
As expected, these are positive-definite quantities.

This is of course not the whole housekeeping entropy.
The remaining part is denoted by $\Delta S^\sigma_{\rm as}=\Delta S_{\rm hk}-\Delta S^\sigma_{\rm{bDB}}$.
From the relation
\begin{equation}
\Delta S_{\rm env}=\Delta S_1 + \Delta S^\sigma_{\rm{bDB}} +\Delta S^\sigma_{\rm as},
\label{S_sum}
\end{equation}
and from Eqs.~(\ref{dS_env}), (\ref{dS_1}), and (\ref{dS_h}),
we obtain
\begin{equation}
dS^\sigma_{\rm as}=
\ln\left[\frac{\Gamma^\ast[\bm{q},t+dt|\bm{q}^\prime,t]}{\Gamma[\bm{q}^\prime,t+dt|\bm{q},t]}
\frac
{\Gamma^\ddag_{h_\sigma}[\bm{q}^\prime,t+dt| \bm{q},t]}
{\Gamma[\bm{\epsilon q},t+dt|\bm{\epsilon q}^\prime,t]}\right].
\end{equation}
This cannot be expressed as the log-ratio of the two conditional probabilities, and therefore
there is no IFT for
$\Delta S^\sigma_{\rm as}$.
Using Eqs.~(\ref{dS_env2}), (\ref{dS_11}), (\ref{dS_sigma}) and
(\ref{S_sum}), we have
\begin{eqnarray}
&&d S^\sigma_{\rm as}=dt\Big[ \dot{q}_i \partial_i \phi_{\rm A}(\bar{\bm{q}})
+\sigma \partial_i \{D_{ij}(\bar{\bm{q}})\partial_j \phi_{\rm A}(\bar{\bm{q}})\} \Big]
\nonumber \\
&&~~~~-\sigma dt (\partial_i \phi_{\rm A}(\bar{\bm{q}}))\Big[ \dot{q}_i
+\epsilon_i A_i ( \bm{ \epsilon} \bar{\bm{q}})-\partial_j D_{ij}(\bar{\bm{q}})
 \nonumber \\
&&~~~~~~~~~+2D_{ij}(\bar{\bm{q}})\partial_j \phi(\bm{\epsilon} \bar{\bm{q}})
+\sigma D_{ij}(\bar{\bm{q}})\partial_j \phi_{\rm A}(\bar{\bm{q}}) \Big].
\end{eqnarray}
We can see that this expression
vanishes when the stationary state distribution is symmetric, i.e. $\phi_{\rm A}=0$.
Therefore, $\Delta S^\sigma_{\rm as}$ can be regarded as the EP due to the asymmetry of
the stationary state distribution. We also note that when $\sigma=0$, we reproduce the similar one in
Ref.~\cite{spinney} (called $dS_3$ there).
The average rate of this EP can be obtained as before. We obtain
\begin{eqnarray}
&&\left\langle \frac{d S^\sigma_{\rm as}}{dt} \right\rangle
=\int d\bm{q}\; \Big[\phi_{\rm A}(\bm{q}) \partial_t\rho(\bm{q})
\label{dS_as_rate}\\
&& ~~ -\sigma(\partial_i \phi_{\rm A}(\bm{q}))\Big\{
\frac{2\epsilon_i j_i ^{\rm s,ir}(\bm{\epsilon q})}{\rho^{\rm s}(\bm{\epsilon q})}
 +\sigma D_{ij}(\bm{q})\partial_j \phi_{\rm A}(\bm{q})\Big\}\rho(\bm{q})\Big].
 \nonumber
\end{eqnarray}
This quantity is obviously not positive definite as there is no IFT for $\Delta S^\sigma_{\rm as}$.
The first term is exactly $\langle d S_3/dt \rangle $ obtained by Spinney and Ford \cite{spinney},
which is a transient contribution to the EP. For any nonzero $\sigma$, however, this EP is not transient and can be regarded as a relevant part of the adiabatic housekeeping EP.

As an example, we consider the one dimensional system driven by a constant force $F$ and a momentum-dependent force $-G'p/m$. We have $\dot{x}=p/m$ and
\begin{equation}
 \dot{p}=-\frac{G}{m}p-\frac{G'}{m}p+F+\xi(t)
\end{equation}
with $\langle\xi(t)\xi(t^\prime)\rangle=2D\delta(t-t^\prime)$ and $D=GT$.
Suppose that the system has reached the stationary state described by
the distribution \cite{spinney}
\begin{equation}
 \rho^{\rm s}(p)=\sqrt{\frac{1}{2\pi mT_{\rm eff}}}\exp\Big[ -\frac{(p-\langle p \rangle_{\rm s})^2}{2mT_{\rm eff}}\Big],
\end{equation}
where $T_{\rm eff}=GT/(G+G')$ and
\begin{equation}
\langle p\rangle_{\rm s} =\frac{mF}{G+G'}.
\end{equation}
In this case, we have
\begin{equation}
 \phi_{\rm A}(p)= -\frac{2\langle p \rangle_{\rm s}p}{mT_{\rm eff}},
\label{examp_1}
\end{equation}
and therefore
$ D\partial_p  \phi_{\rm A}(p)=-2(G+G')\langle p \rangle_{\rm s}/m$.

The combination in Eq.~(\ref{dS_sigma_rate}) is given by
\begin{equation}
 \frac{j^{\rm s,ir}_p(-p)}{\rho^{\rm s}(-p)}=\frac{(G+G')p}{m}-D\partial_p\phi(-p)=-\frac{(G+G')\langle p \rangle_{\rm s}}{m}.
\label {examp_2}
 \end{equation}
Inserting Eq.~(\ref{examp_1}) into Eq.~(\ref{dS_sigma_rate}), we have
the EP rate in the steady state as
\begin{equation}
 \left\langle \frac{d S_{\rm bDB}^\sigma}{dt}  \right\rangle_{\rm s}=(2\sigma-1)^2\frac{(G+G')^2\langle p\rangle^2_{\rm s}}{Dm^2}
 =(2\sigma -1)^2 \frac{F^2}{D},
\end{equation}
which is always positive, as expected from the IFT.
Recall that we are considering the case where $\sigma\neq 1/2$.
From Eq.~(\ref{dS_as_rate}), we have
\begin{equation}
 \left\langle \frac{d S^\sigma_{\rm as}}{dt}  \right\rangle_{\rm s}=-4\sigma(\sigma-1) \frac{F^2}{D}.
\end{equation}
Then $\langle dS_{\rm hk}/dt\rangle_{\rm s} =\langle d(S_{\rm bDB}^\sigma+S^\sigma_{\rm as})/dt\rangle_{\rm s}=F^2/D$, which is independent of $\sigma$.

\section{Summary and Discussion}

In summary, we have shown how the housekeeping EP $\Delta S_{\rm hk}$ for the continuous stochastic
dynamics in the presence of odd-parity variables
can be separated into two parts with the former satisfying the IFT. As we have seen,
the presence of the odd-parity variables and the possibility of the asymmetry
of the stationary state distribution makes this separation nontrivial.  For the case of discrete state variables, one particular quantity
$\Delta S_{\rm bDB}$ stands out, which
has the physical meaning of describing the departure from the DB~\cite{lkp}.
In the continuous variable cases, however, the corresponding quantity turns out to be ill-defined.
By considering a generalized adjoint process, we have shown that there is one-parameter family of the EPs for a range of values of the
parameter $\sigma$ belonging to $\Delta S_{\rm hk}$.
The previously known EPs in Ref.~\cite{spinney} are included as a special case for $\sigma=0$.

We have shown that the DB condition is equivalent to the nonvanishing irreversible current and the parity symmetry in the stationary state, $\mathbf{j}^{\textrm{s,ir}}=0$ and $\phi_{\textrm{A}}=0$. We exploited the adjoint process in Eq.~(\ref{omega_ddag}) parametrized by Eq.~(\ref{h_sigma}) to extract an EP satisfying the IFT from the log-ratio of the two path probabilities for the adjoint process and the original process, respectively. We found the two processes to be equivalent if the DB condition is satisfied. Then, the obtained EP $\Delta S_{\rm bDB}^\sigma$ is directly related with the breakage of the DB, characterized by both nonzero $\mathbf{j}^{\textrm{s,ir}}$ and the parity asymmetry $\phi_{\textrm{A}}\neq 0$, as seen in Eq.~(\ref{dS_sigma_rate_1}). The remaining part $\Delta S^\sigma_{\rm as}$ not satisfying the IFT is also responsible for the breakage of the DB solely due to $\phi_{\textrm{A}}\neq 0$, as seen in Eq.~(\ref{dS_as_rate}). 

In this paper, we have also investigated the similarities and differences between the discrete jumping processes and the continuous stochastic processes described by
the master equation and the Fokker-Planck equation, respectively.
It may not be surprising that some expressions obtained in the discrete variable case cannot be directly translated into
the continuous variable case. In fact, the DB condition automatically implies the parity symmetry of the stationary state
distribution for the continuous variable case. 
When the variables are discrete, however, the two conditions are independent. 
We have found that this difference in the DB condition is responsible for the degeneracy in $\Delta S_{\rm bDB}^\sigma$
for the continuous variable case.
In the discrete jumping process, the remaining part of the housekeeping EP apart from the one associated with
the breakage of DB has been shown to be not transient~\cite{lkp}. This term measures the parity asymmetry of the stationary state
distribution.
We also note that the corresponding quantity in our case,
$\Delta S^\sigma_{\rm as}$, is also not transient for nonzero values of the parameter $\sigma$. It is consistent with the persistency of the nonequilibrium steady state with the broken DB. Only when $\sigma=0$, this contribution exists only transiently, which corresponds to the case 
studied by Spinney and Ford \cite{spinney}.

\begin{acknowledgments}
This research was supported by the NRF Grant Nos.~2014R1A1A2053362 (J.Y.), 2013R1A1A2011079 (C.K.),
2014R1A3A2069005 (H.K.L.), and 2013R1A1A2A10009722 (H.P.).
\end{acknowledgments}

\end{document}